# Appropriation of social media by patients with chronic illness to share knowledge

**Nwakego Ugochi Isika**
Department of Computing and Information Systems
University of Melbourne
Melbourne, Australia
Email: nisika@student.unimelb.edu.au

**Antonette Mendoza**
Department of Computing and Information Systems
University of Melbourne
Melbourne, Australia
Email: mendozaa@unimelb.edu.au

**Rachelle Bosua**
Department of Computing and Information Systems
University of Melbourne
Melbourne, Australia
Email: rachelle.bosua@unimelb.edu.au

## Abstract

Social media technologies are increasingly utilized by patients, leading to development of online social groups where patients share experiences and offer support to their peers on these platforms. Yet, there is limited research investigating actual use of: social media platforms by patients; issues faced in using such platforms and how appropriation of these platforms impact patient outcomes. In this study, we propose a conceptual model that encapsulates Social Support Theory and Model of Technology Appropriation to investigate the role of social media and support networks in the health care domain, especially for patients suffering from chronic illnesses. We propose that Social Support Theory and the Model of Technology Appropriation could explain the determining factors both the enablers and barriers that drive appropriation and knowledge sharing behaviours of patients on social media platforms.

## 1　INTRODUCTION

The role of patients in healthcare has recently changed significantly with patients being expected to actively participate in their care and treatment (Adams, 2010). This change is even more significant with the uptake of social technologies that mediate this participation. To this effect, social media platforms are increasingly utilized in the healthcare domain by patients with chronic illnesses (Househ et al, 2014). Social groups formed online, enable patients with rare illnesses to find others beyond the geographical barriers that once isolated sufferers of chronic illnesses (Fourie & Julien, 2014). In addition, patients armed with information from their cohorts are empowered to take a more active role in their personal health management. Furthermore, these platforms are not only taken up by patients but also by family members and friends to be better informed on the situation the patient is going through. Previous research has noted that increased usage of this platform by patients with chronic illnesses provide patients the following: emotional support, ability to seek information from people in similar circumstances, a forum to share knowledge and experiences as well as a plethora of other benefits ( Lober & Flowers, 2011; Rozenblum & Bates, 2013; Househ et al., 2014). The majority of the available studies are focused on predecessors of social media such as discussion forums, bulletin boards and so on (Merolli et al, 2013). Prior studies suggest that little is known about the type of information provided on this platform and how patients participate to make decisions on whether to use or not to use information provided by their peers (Househ et al., 2014; Fernández-Luque & Bau, 2015) . These studies assert that there is a lack of in depth knowledge about how people use this technology, and concur that understanding this would lead to patient empowerment and improved health outcomes. Particularly, there is limited knowledge on the use of social media for health, and call for researchers from multiple disciplines to examine this phenomena. Furthermore, the majority of the literature examined has been from an intervention-based or medical view-point. Acquiring an understanding of how this platform is appropriated for use by patients will be beneficial for health authorities enabling them to leverage these tools in supporting public health efforts (Fernández-Luque & Bau, 2015).

Hence the guiding question to be addressed by this study is: *How do patients appropriate social media sites such as Facebook.com to share knowledge with their cohorts?* Moreover, this investigation also explores the role of emotion on appropriation in this context. A growing body of IS literature notes that the decision to adopt and use a technology is often driven by the user's emotion (Agarwal et al, 2010; Mendoza et al, 2013).

We define appropriation as the relationship between the way the user desires to utilize a system, the capabilities of that system and the situations in which these systems are used (Carroll et al., 2002). Appropriation is seen in IS literature as a subjective process, where the meaning and use of technology is dependent on context and technology is transformed through the utilization process (Mackay & Gillespie, 1992).

Further, we define social media platforms as a set of internet based tools that facilitate creation and exchange of user generated content (Ngai et al., 2015). These platforms include technologies such as blogs, video-sharing sites, micro-blogs, forums, websites and wikis that are developed based on the concepts of Web 2.0 ( Preece, 2001; Househ et al., 2014; Ngai et al., 2015).

In order to fully explore this phenomenon, a qualitative case study methodology is to be applied in this investigation. Currently, a pilot study is being conducted on a Reddit subreddit for patients living with fibromyalgia to gain a deeper understanding of this issue. Reddit.com is a social media site that caters to users with diverse interests. The system allows users create virtual rooms or "subreddits" according to their interests. Other users are able to participate in this room and site appointed moderators control the content in these subreddits to ensure they comply with the code of conduct stated for that subreddit. The next step in this investigation is to conduct two case studies of two other platforms with patients managing other chronic illnesses such as prostate cancer or diabetes. Data collection methods for the investigation will consist of content analysis of data obtained from these sites which will be open coded and then thematically coded and a questionnaire to fully examine this phenomenon using the Model of Technology Appropriation and Social Support Theory.

## 2   THEORETICAL BACKGROUND

Online communities, social media, virtual communities and online social groups are used interchangeably in the social media literature in this domain (Dungay et al., 2015). In this study, social media refers to a phenomenon where groups of people, bound by a shared motivation, come together virtually for a joint enterprise (Wenger, 2004). Social media platforms as an IS research environment has a number of challenges such as: the ethical implications of utilizing user data located online to conduct studies, as well as debate on the appropriate methodologies with which to analyse and understand data (Agarwal et al, 2010). However, social media platforms also provide IS researchers with a new environment in which to carry out qualitative research, and the availability of content for quantitative, qualitative and tool based analysis ( Vannoy & Palvia, 2010; Pousti et al., 2013; Cao et al, 2015). Agarwal et al (2010) address the digitization of healthcare and examine the opportunities for IS research in healthcare organizational context. Additionally, they highlight non-traditional areas such as social media appropriation by patients with the advent of patient-centred virtual communities, which provide opportunities for IS scholars in this domain (Agarwal et al, 2010). Furthermore, the study indicates that issues related to adoption and use of social media are potential research areas for IS scholars. Consequently, examination of tools utilized by patients, patterns of use of these tools, effects of appropriation on patient's health outcomes and their potential impacts on physician-patient relationships are fruitful research opportunities in the IS domain (Agarwal et al, 2010).

Typically, IS research on system appropriation take the traditional viewpoints of ease of use and usefulness as illustrated in theories such as Theory of Reasoned Action (TRA), Technology Acceptance Model (TAM), Theory of Planned Behaviour (TPB), Model of Technology Appropriation (MTA) and traditional technology use and adoption theories. However in a social media environment a key function is the social aspect of these platforms which begs the addition of theories explaining human behavioural influences in these environments (Vannoy & Palvia, 2010).

It seems that user personalization and an emphasis on sociability are core components of social media today, with the ability to create online profiles and communicate asynchronously with peers (Merolli et al., 2013). Additionally, appropriation of such social media is much easier than standalone tools because these technologies are accessible, low-cost and easy to use. Use of websites such as 'patientslikeme' (https://www.patientslikeme.com) are used to enable sharing of medical records with peers, leading to improved disease management outcomes and emotional benefits for participants (Adams, 2010; Househ et al., 2014).

Virtual communities of patients with chronic illness differ from educational or organizational virtual communities in terms of using emotional language and intolerance for harsh or negative communication, while social support is an essential facet of these communities ( Preece, 2001; Bender et al., 2013; Pousti et al., 2013). Bender et al. (2013) investigates breast cancer support groups in their study that followed a mixed-methods approach using interviews and surveys. Bender et al (2013) also examined factors influencing patients to seek online support for illnesses. The authors indicate that rareness of a condition, embarrassment, need for support, need for information and fear of stigmatization cause a patient to seek support online.

The majority of participants in these groups discovered each other while searching for information on their conditions online. Motivations for using social media groups included the desire for support and information from people suffering the same conditions as other participants of the study (Bender et al., 2013). Most of the study participants had adopted and then discarded other groups for reasons such as: anxiety due to insensitive information, lack of engagement by other users and because information that was provided to them was not relevant to their situation. An interesting insight from this study was the recommendation to use a multi-theory lens to examine this phenomenon because of the emotional and psychological factors that are in play when patients with chronic illness perform this behaviour (Bender et al., 2013).

However, the population studied in this paper were breast cancer survivors who were facilitating face-to-face social support groups. Further, the methods used for the study were qualitative interviews and surveys, which could create a self-reporting bias from the respondents and yield different results from what could be found through a different method that examines the actual phenomenon in its context. Additionally, the study was conducted retrospectively by patients who had survived their illness and who were now facilitators of face-to-face breast cancer groups and could therefore be prone to recall and self-reporting bias (Bender et al., 2013).

Merrolli et al (2013) conducted a literature review to discover affordances, health outcomes and effects of social media use in chronic disease management. Social support and improved psychosocial outcomes were discovered as a benefit of participation in these communities. In addition, a measure of control over their identities, access to information for disease management, flexibility afforded through asynchronous communication with peers, ability to share experiences and the flexibility of the system were also reported as benefits to participation on this medium (Merolli et al., 2013). Merolli (2013) recommends further studies to gain a deeper understanding of the use of social media by this group. Their studies suggest that social media could be an effective tool to support the management of chronic illness both by patients and from an organizational perspective. Some previous studies indicate that Social Support Theory is a sufficient lens for examining knowledge sharing and emotional aspects of the use of this platform (Bender et al., 2013; Guo & Goh, 2014; Huh et al., 2014; Fernández-Luque & Bau, 2015; Huh et al., 2015). However, studies in this area seem to take an intervention or information behaviour approach that does not shed much light on the mediating role of technology.

The next section discusses some technology adoption and use theories commonly used to investigate appropriation in IS research. These seminal studies are used in the information system domain to develop insights into the interaction between users and technology.

## 2.1 Technology adoption and use

Appropriation refers to how a user *adopts* and *uses* a technology. Orlikowski (1999) terms actual technology use as "technology-in-practice", as these processes could be faithful to the intended use of the technology designers or could diverge based on user preferences or understandings on how the technology works (DeSanctis & Poole, 1994; Orlikowski, 1999). In addition, decision to use these technologies are dependent on user motivations (Orlikowski, 1999). Appropriation of a system is context-specific as a number of factors influence the decision to adopt and how the system is used after it has been adopted (Orlikowski, 1999; Orlikowski, 2009).

Fishbein & Ajzen (1975) in their proposal of the Theory of Reasoned Action (TRA), suggest that individuals' intention to engage in any behaviour is the moderating factor in predicting their behaviour. Key concepts of this theory include the supposition that behavioural intention is determined by the individual's attitude towards a behaviour and subjective norm. A key drawback of this seminal theory is 'intention' as the moderator, as intention can only be applied for analysis of voluntary behaviour (Conner & Armitage, 1998; Fishbein & Ajzen, 1975). Therefore, the authors extended their theory to include *involuntary behaviour* in The Theory of Planned Behaviour (Ajzen & Madden, 1986). This theory (TPB) posited by Ajzen (1991), expands on the theory of reasoned action proposed by Fishbein & Ajzen (1975) with the focus on predicting user behaviour on computer systems. Decisions to undertake behaviour by individuals is considered a result of careful consideration of available information concerning participation in that behaviour(Ajzen & Madden, 1986; Ajzen, 1991). TPB can be successfully applied to voluntary and involuntary behaviour of individuals as it incorporates control perceptions as a moderator for behaviour outcome. Similarly, the Technology Acceptance Model (TAM) illustrates that perceptions, intentions and attitudes to the system, predict adoption and use (Davis, Bagozzi, & Warshaw, 1989; Venkatesh et al, 2003). Venkatesh et al (2003) extend TAM to include social factors as influencers in the technology appropriation process, by proposing a unified theory that combines eight seminal theories in IS adoption and use to explain appropriation of software systems. The model developed was the Unified Theory of Acceptance and Use of Technology (UTAUT).

Likewise, Structuration Theory has been employed in IS field by researchers who proposed that users shape technology through their usage contexts (Giddens, 1984; Orlikowski, 1992). This theory was extended by Desanctis & Poole (1994) who proposed Adaptive Structuration Theory (AST) as a means of investigating appropriation in terms of two view-points: structures built into technology and emergent structures that develop from human interaction with these technologies (DeSanctis & Poole, 1994). AST was found to be too rigid to explain adaptation of technology over time. Therefore it was suggested that an organization element be taken into account to ensure adaptation could be fully explained by this theory (Majchrzak et al., 2000).

The Model of Technology Appropriation (MTA) is also used to examine technology-in-use, and defines appropriation as the relationship between the way the user desires to utilize a system, the capabilities of that system and the situations in which these systems are used (Carroll et al, 2002). Mendoza et al, (2008) examine long-term use of a system in an academic setting using MTA, and discovered that the appropriation process is incremental and the definitions of usefulness change over time as users become adept in utilizing the system. In addition social norms, institutional mandates and support mechanisms were found to be key influencers in appropriation. Findings indicate that subjective norm(s), ease of use, usefulness and training impact appropriation in this domain (Carroll et al., 2002; Mendoza et al., 2008; 2010). Little attention is paid to understanding the influence of emotional aspects on appropriation of a system. However, the model is quite equipped to examine the phases from initial system adoption to longer term use of a system.

Recent studies in IS indicate that emotion should be accounted for when evaluating appropriation of a system. Emotion and affect are used interchangeably in IS literature and play a pivotal role in determining the information behaviour of a user, (Fourie & Julien, 2014). Mendoza et al (2013) examine the role of emotions and quality needs on appropriation of a system by users. Their study revealed that emotions play a role in the appropriation process despite the fact that they examined two very different socio-technical systems: a learning management system and a personal alarm system.

The following section discusses the concepts of knowledge and knowledge sharing as defined in this study.

## 2.2   Knowledge sharing and emotional support on social media platforms

Knowledge is defined in this context as the experiences and understanding of an individual which is then translated into a meaningful form (i.e. an action) that is applied by the individual to carry out a task (Panahi et al., 2012).

Knowledge sharing is typically defined in an organizational context, however this study defines knowledge sharing as the communication of knowledge from source to recipient, such that the recipient is able to utilize this knowledge in their own activities and thinking (Ma & Chan, 2014; Panahi et al., 2012). Panahi et al (2012) state that knowledge can be transferred through storytelling and narratives and technology acts as an enabler. The success of extended online participation in virtual groups depends on continuous, meaningful knowledge sharing by its members ( Hsu et al., 2007; Fang & Chiu, 2010; Belous, 2014).

Additionally the success of an online environment is often dependent on the exchange of knowledge between the seeker and the provider.

## 2.3   Social theories

In his seminal article Emerson (1976) posits that relationships developed by individuals are for mutual tangible or intangible benefits and termed this phenomena social exchange theory (Emerson, 1976). The concept behind this theory is that social exchange is contingent on actions motivated by rewarding reactions of others within a social group, implicitly mutually rewarding to all individuals within this network. This school of thought acknowledges that for exchange behaviour to occur, external and internal factors play influential roles either encouraging or prohibiting this behaviour (Baldwin, 1978).

Likewise, Social Cognitive Theory has been widely used in IS research as it examines human behaviour as a result of the interplay between personality, social networks and behaviour (Chiu et al., 2006). Bandura postulated this theory with two key moderators for behavioural outcomes: self-efficacy and outcome expectations (Bandura, 1989). Previous studies confirm this viewpoint revealing that self-efficacy and outcome expectations such intrinsic rewards impact knowledge sharing in online communities such as a recreation-oriented interest groups (Lai & Chen, 2014).

Similarly, Social Capital Theory is commonly used to examine social network interactions. This key ideology of this theory indicates that social ties and relationships are productive resources which can be harnessed by members of a network for mutual benefit (Chiu et al., 2006). Also, Social Capital Theory posits that structural, cognitive and relationship social capital predicts user behaviour (Nahapiet & Ghoshal, 1998). In addition, Nahapiet & Ghosal (1998) suggest that social capital requires an investment that is related to the size of the network and technology could play a role in building Social Capital.

A key theory for examining social interactions in the healthcare domain is Social Support Theory. Shumaker & Brownell (1984) define social support as "*an exchange between two or more parties in which the provider or the recipient perceives the exchange to be beneficial to the recipient. There are costs and benefits to participants in this activity, with mental and physical implications in the absence or presence of stressors* " (Shumaker & Brownell, 1984).

A key feature of this theory is the concept of *reciprocity*. Shumaker & Brownell (1984) state that an example could be seen in mutual care-giving in self-help groups as opposed to the common viewpoint that reciprocity goes from the recipient to the provider of support.

A recurring theme in the literature examined in the theoretical background section is a need for provision of emotional support by people with a shared medical condition (Greene et al., 2011, Bender et al., 2013; Merolli et al., 2013; Househ et al., 2014; Dungay et al., 2015). Fourie & Julien (2014) suggest researchers take note of emotional aspects involved in human behaviour as pertinent to IS research. Particularly in the circumstance where patients on these platforms are dealing with an illness that has emotional implications to them (Bender, Katz, et al., 2013; Merolli et al., 2013; Fourie & Julien, 2014). In situations where users require support and need to cope with stressful consequences of their ailment, these emotions could have an impact in their motivation and actions - in this case the decision to participate in these communities and how they participate in these communities (Bender, Katz, et al., 2013; Pousti et al., 2013).

Similar to the other social theories discussed above, Social Support Theory (SST) hinges on pro-social behaviour, which is influenced by characteristics of the provider and recipient in this exchange. Uniquely, Social Support Theory could also explain the behaviour of patients in this environment for instance, willingness to disclose personal information to peers online ( Shumaker & Brownell, 1984; Huh et al, 2014; Rubenstein, 2015). Social support promotes a sense of virtual community by offering a buffer for members against the effects of stress, leading to continuance behaviour (Huh et al., 2014; Rubenstein, 2015). In addition, previous studies found that a lower amount of offline support correlated with higher engagement with social media by patients.

In order to the address the limitations mentioned above, we introduce a conceptual model that underpins this study. This section discusses the use of social support theory as a lens for examining appropriation in this investigation.

## 3   CONCEPTUAL MODEL

We propose a conceptual model (Figure 1) that integrates Social Support Theory and the Model of Technology Appropriation to interpret the initial findings from the study. Using Social Support

Theory in combination with the Model of Technology Appropriation as theoretical lens, this study examines: (1) the appropriation process undertaken by patients using social media and (2) identify the factors influencing appropriation of social media by patients with chronic illness for knowledge sharing. The research question for the study is: *How do patients appropriate social media sites such as Facebook.com to share knowledge with their cohorts*. To gain a deeper understanding into this phenomenon, we aim to examine the relationships between people's emotions and information behaviour while using social media platforms. The conceptual model in Figure 1 combining the Model of Technology Appropriation with Social Support Theory could provide insights to this study. Social Support Theory is thought to have three components: informational support, emotional support and social support (Rubenstein, 2015). Social support theory could explain the issue under examination while being cognizant of the emotional needs of participants in this environment. Social support is thought to comprise emotional, social and informational dimensions, while the model of technology appropriation is concerned with adoption and long-term use of systems.

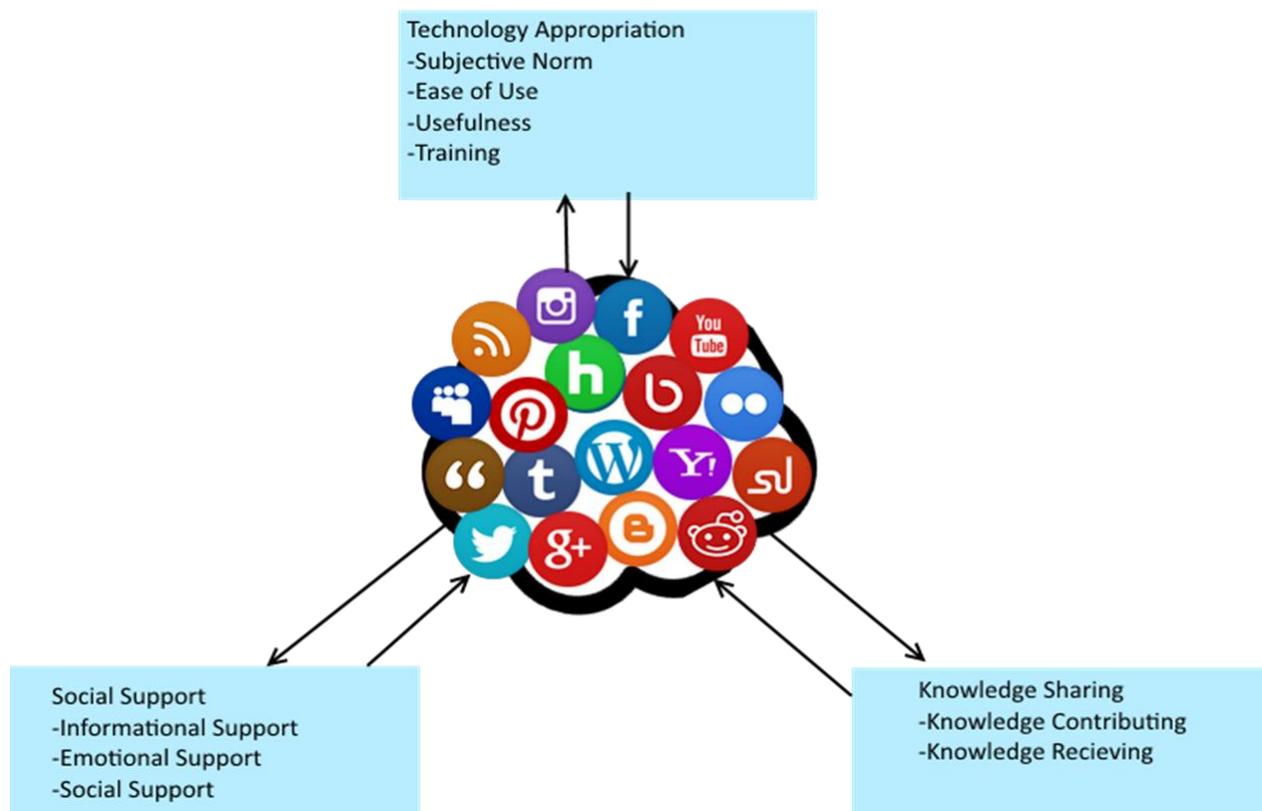

*Figure 1: Conceptual research model for the study*

The viewpoint taken on these theories as illustrated above is that knowledge sharing activities, social support and availability of appropriation factors have a symbiotic relationship in initial adoption phase and long-term use of social media platforms. The conceptual model above outlines the knowledge sharing and receiving activities as contributors to the perceived social support for patients on these platforms. Combining these two theories enables the study to take the emotional aspect of this phenomenon into account while examining appropriation. Therefore an initial conceptual model is proposed as a basis for this study in Figure 1 combining social support, appropriation and knowledge sharing ideologies. The next section discusses the proposed research methodology of the study.

## 4    RESEARCH METHODOLOGY

The study will follow a qualitative, interpretive case study method in order to examine appropriation of social media platforms for knowledge sharing by patients with chronic illness. The intention of an interpretive case study is to understand key phenomena-in- context taking into account participants' views (Cavaye, 1996; Andrade, 2009). *There is no, and never will be, a best theory. Theory is our chronic inadequate attempt to come to terms with the infinite complexity of the real world. Our quest* *should be for improved theory, not best theory, and for theory that is relevant to the issues of our time* (Walsham 1993 p. 478).

The theoretical foundations discussed above provide a scaffold for the creation of a model through which this phenomena can be explained (Dungay et al., 2015). Case studies are used to handle research dealing in *how or why*-type questions (Yin, 2013). The qualitative approach in this study will allow the discovery of rich experiences of participants by the researchers.

### 4.1 Research Aims

The research aims are derived from limitations identified in the literature review, which will be ongoing throughout the study. Specifically this study aims to:

- Understand how patients with chronic illnesses appropriate online social platforms to share knowledge
- Establish an understanding of why patients appropriate online social platforms for knowledge sharing activities
- Understand user participation over time on online social platforms
- Understand the role of emotions in appropriation of social media for knowledge sharing by patients with chronic illness.

In order to achieve the aims listed above, the following research questions serve as a guide to this enquiry.

### 4.2 Research Questions

The study aims to answer the following research questions:

- How do patients appropriate social media platforms to share knowledge?
- Why do patients participate in knowledge sharing on such social media platforms?
- What factors influence their continued participation in social media platforms for knowledge sharing?
- How does *emotion* impact appropriation of social media platforms for knowledge sharing by patients with chronic illness?

The proposed research will consist of two case studies of online environments such as dailystrength.com and inspire.com as they fit the criteria defined for the study. The approach enables the identification of themes, similarities and disparities between cases. Therefore, a model could be developed based on the findings. A pilot study is currently being undertaken using content retrieved from a Reddit subreddit thread for patients living with fibromyalgia. All activities for the subreddit were collected for analysis with the intention of identifying key themes to drive the next steps of the research.

Following the pilot study, two case studies on inspire.com and daily strength for patients with another chronic illness will be examined. These sites are chosen because the structure of the systems is such that the entirety of the user's activities on the site is available in a linear digital trail to indicate their action from joining the site until the date they last engaged with the platform. This data will be collected for analysis and anonymized with permission from the site providers. This enables the authors to conduct a longitudinal study to understand this appropriation and social support phenomenon. Subsequently, a questionnaire instrument will be used to collect data concerning the technology adoption, use and social support factors.

The study intends to examine the interaction on these platforms to generate a model describing the phenomenon. The study will generate the outcomes to the chosen theories based on findings for theoretical replications (Yin, 2013). Also, multiple case studies will ensure replicability and minimize the potential impact of extreme cases on findings (Yin, 2013).

## 5　POTENTIAL CONTRIBUTIONS AND CONCLUSION

With the advent of social media platforms, patients are increasingly utilizing online social groups to share experiences and offer support to their peers on these networks. Our study aims at examining the appropriation of such social media platforms to examine knowledge sharing by patients with chronic illnesses and how technologies act as enablers or barriers to this process. This research-in-progress study is intended to provide a holistic model that encapsulates appropriation of socio-technical systems by patients with chronic illness. The proposed model would enable a deeper understanding of the factors that impact this phenomenon, while also shedding light on the perceived social support afforded to the user through interaction with these systems in the healthcare domain.

Future work will include: (1) conducting case studies on a variety of social media platforms used by patients suffering from chronic illnesses; (2) the development of appropriation behaviours in the healthcare domain and; (3) design of appropriation metrics on how best to evaluate emotional impacts of social platforms for patient-centred healthcare. Furthermore, our findings will contribute to the body of knowledge in the medical/IS and KM domains allowing contribution to social media intervention-based research and service providers of such domains.

## 6　REFERENCES

## Copyright